\documentclass[showpacs,preprintnumbers,amsmath,amssymb,APSl,prd,nofootinbib,superscriptaddress, onecolumn]{revtex4}
\usepackage[dvips]{graphicx}
\usepackage{amssymb}
\usepackage{amsmath}
\usepackage{graphicx,color}
\usepackage{amsfonts}
\usepackage{bm}
\usepackage{cancel}
\usepackage{comment}
\usepackage{hyperref}
\usepackage{babel}

\newcommand{\be}{\begin{equation}}
\newcommand{\ee}{\end{equation}}
\newcommand{\bea}{\begin{eqnarray}}
\newcommand{\eea}{\end{eqnarray}}
\newcommand{\beaa}{\begin{eqnarray*}}
\newcommand{\eeaa}{\end{eqnarray*}}

\allowdisplaybreaks[4]

\begin{document}

\tolerance=5000

\title{Exponential $F(R)$  gravity  with axion dark matter}

\author{Sergei D. Odintsov}
\email{odintsov@ice.csic.es} \affiliation{Institute of Space Sciences (ICE, CSIC) C. Can Magrans s/n, 08193 Barcelona, Spain}
 \affiliation{Instituci\'o Catalana de Recerca i Estudis Avan\c{c}ats (ICREA),
Passeig Luis Companys, 23, 08010 Barcelona, Spain}
 \author{Diego~S\'aez-Chill\'on~G\'omez}
\email{diego.saez@uva.es}
\affiliation{Department of Theoretical, Atomic and Optical
Physics and IMUVA, Campus Miguel Delibes, \\ University of Valladolid UVA, Paseo Bel\'en, 7, 47011
Valladolid, Spain}
\author{German~S.~Sharov}
 \email{sharov.gs@tversu.ru}
 \affiliation{Tver state university, Sadovyj per. 35, 170002 Tver, Russia}
 \affiliation{International Laboratory for Theoretical Cosmology,
Tomsk State University of Control Systems and Radioelectronics (TUSUR), 634050 Tomsk,
Russia}

\begin{abstract}
The cosmological evolution within the framework of exponential $F(R)$ gravity is analysed by assuming two forms for dark matter: (a) a standard dust-like fluid and (b) an axion scalar field. As shown in previous literature, an axion-like field oscillates during the cosmological evolution but can play the role of dark matter when approaching the minimum of its potential. Both scenarios are confronted with recent observational data
including the Pantheon Type Ia supernovae,  Hubble parameter estimations (Cosmic Chronometers), Baryon Acoustic Oscillations and Cosmic Microwave Background distances. The models show great possibilities in describing these observations when compared with the $\Lambda$CDM model, supporting the viability of exponential $F(R)$ gravity. The differences between both descriptions of dark matter is analysed.
\end{abstract}
%
%
%\pacs{04.50.Kd, 98.80.-k, 95.36.+x}
%%
\maketitle

\section{Introduction}\label{Intr}

Extensions of Einstein's General Relativity (GR) have been widely analysed in the literature over the last two decades, motivated mainly to shed some light on one of the main challenges of theoretical physics for this twenty first century: the explanation of the cosmological evolution and particularly the late-time acceleration of the universe expansion, or in other words, the nature of the so-called dark energy. This is based on the assumption that GR fails to explain the universe at large scales, where some corrections to Einstein's theory might play an important role (for a review on extensions of GR see Refs.~\cite{Nojiri:2010wj}). The simplest way of extending GR consists on adding a cosmological constant to the Hilbert-Einstein action, motivated also somehow on the possibility of the gravitational effects of the vacuum energy density as provided by quantum field theories. This is the so-called $\Lambda$ Cold Dark Matter model ($\Lambda$CDM), which besides the cosmological constant that provides an explanation to late-time acceleration, the model includes the so-called dark matter in the form of a pressureless fluid that turns out  fundamental for the formation of large scale structure, among other stuff. As shown by the observational data, the standard model of cosmology has been incredibly successful but not without important issues (for a review on the state of the art of current cosmology in comparison to observational data, see \cite{DiValentino:2020vhf}).  In addition, as far as one assumes classical gravitation as an effective theory, some other corrections beyond might arise in the gravitational action. In this sense, the next step would be to include non-linear terms of the Ricci scalar in the action, the so-called $F(R)$ gravities. As shown in the literature (see Refs.~\cite{Capozziello:2002rd}), $F(R)$ gravities have become very popular and interesting as can give rise to any cosmological solution by choosing the appropriate action. In this sense, $F(R)$ gravities can describe not only late-time acceleration but also the inflationary phase at the beginning of the universe evolution, as shown by some of the most promising models of inflation according to their predictions in comparison to the Planck data \cite{Starobinsky:1980te}.  Moreover, one might construct the appropriate $F(R)$ action that unifies late-time acceleration and inflation, where both arise as a consequence of the presence of non-linear terms of the Ricci scalar in the action \cite{Nojiri:2007as,Nojiri:2005pu}. In addition, such models seem to be capable to recover GR at some local limits, turning out a viable possibility and satisfying the observational constraints \cite{Nojiri:2007as,HuSawicki07,delaCruz-Dombriz:2015tye}. \\

On the other hand, dark matter seems to remain slippery for the particle physicists, as no signal of a dark matter particle has been detected so far in the laboratory (for a review see Ref.~\cite{Arbey:2021gdg}). However, the dark matter gravitational effects are widely probed, from the rotational curves of galaxies to the formation of large scale structure, dark matter is necessary to explain the data. In cosmology the problem is usually avoided by assuming the presence of dark matter as a pressureless fluid without going into details about its nature. However, different descriptions for the nature behind dark matter might lead to different effects on the cosmological evolution. One of the most popular candidates for dark matter is the axion, a theoretical particle postulated to provide a solution to the strong CP problem of quantum chromodynamics but which might be a serious candidate for dark matter as far as the population and its mass, generated about the minimum of a particular potential after the Peccei-Quinnsymmetry is broken, are enough \cite{Marsh:2015xka}. Bounds on the parameters describing axions and a possible conversion into photons have been widely analysed in the literature, leading to the possibility to be detected in the future \cite{Marsh:2017yvc}. Also other constraints have been obtained by using pulsar timing and fast radio burst observations \cite{Caputo:2019tms}, imprints on the Lyman $\alpha$ forest observations \cite{Soda:2017dsu}, through superradiance in some compact objects \cite{Cardoso:2018tly} and by effects produced in the early universe \cite{Oikonomou:2022tux,Oikonomou:2023bah}. Definitely, this type of particles under the name of axions provides the necessary properties for dark matter. \\

In the present paper an exponential $F(R)$ model is considered. This type of $F(R)$ models have been previously analysed in the literature with great success \cite{BambaGL:2010,ElizaldeNOSZ11,OdintsovSGS:2017,OdintsovSGSFlog:2019}. The main action analysed in this paper is given by:
 \begin{equation}
F(R)=R-2\Lambda\big(1-e^{-bR}\big)+F_{\mathrm{inf}}\, ,
 \label{FR1}\end{equation}
 where $\Lambda$, $b$ are constants and the term $F_{\mathrm{inf}}(R)$ is related with the inflationary epoch which becomes negligible at late times. In order to account for dark matter, two different scenarios are considered together with the $F(R)$ model (\ref{FR1}). The first case lies on the usual assumption of dark matter as a pressureless fluid whereas in the second scenario an axion scalar field $\phi$ plays the role of dark matter. The latter has been studied previously, showing its viability for describing the cosmological evolution in $F(R)$ gravity \cite{Oikonomou:2022tux,OdintsovOik_UniAx:2019,OdintsovOik_Axion:2020,Oikonomou_Uni:2021,OikonomouFTR:2023}.
 Both scenarios are compared with observational data from Supernovae Ia (SNe Ia),  Hubble parameter $H(z)$ estimations from differential ages of galaxies or Cosmic Chronometers (CC), observational manifestations of Baryon Acoustic Oscillations (BAO) and Cosmic Microwave Background radiation (CMB). The results are also compared with the predictions of the $\Lambda$CDM model. Our aim is to show the differences when considering dark matter as an effective pressureless fluid and when the axion field plays the role of dark matter, showing the goodness of the fits on comparison to the $\Lambda$CDM model.\\

The paper is organized as follows: section \ref{Models} is devoted to introduce the models and the dynamical system of equations for the exponential $F(R)$ gravity with the standard dark matter and with the axion. Then, the models are tested with SNe Ia, $H(z)$, BAO  and CMB observational data, briefly described in section \ref{Observ}. Results are analyzed in section \ref{Results} and finally the
conclusions are provided in section \ref{conclusions}.

%%%%%%%%%%%%%%%%%%%%%%%%%%%%%%%%%%%%%%%%%%%%%%%%%
\section{$F(R)$ gravity with axion dark matter}
 \label{Models}
 %%%%%%%%%%%%%%%%%%%%%%%%%%%%%%%%%%%%%%%%%%%%%%%%%%%%%%

Let us start by introducing the action for $F(R)$ gravity together with the corresponding Lagrangians for the matter fields
\cite{OdintsovOik_UniAx:2019,OdintsovOik_Axion:2020,Oikonomou_Uni:2021}:
 \begin{equation}
  S =\int d^4x \sqrt{-g}\bigg[ \frac{F(R)}{2\kappa^2}  + {\cal L}_{m} + {\cal L}_{\phi}\bigg]
  ,\qquad{\cal L}_{\phi}=-\frac12\partial^\mu\phi\partial_\mu\phi-V(\phi)\, .
 \label{Act1}\end{equation}
Here $\kappa^2=8\pi G=M_P^{-2}$, where $G$ and $M_P$  are  the Newtonian gravitational constant and the reduced Planck mass respectively, whereas $ {\cal L}_{m}$ is the matter Lagrangian (baryonic matter, radiation...etc) and  ${\cal L}_{\phi}$ describes the axion field. \\

Along this paper, we are considering an exponential $F(R)$ gravity model given by:
 \begin{equation}
 F(R)=R -2\Lambda\big(1-e^{-\beta{\cal R}}\big), \qquad  {\cal R}=\frac{R}{2\Lambda}\ ,
   \label{FR2}
\end{equation}
 where $\beta$ is a dimensionless constant and ${\cal R}$ is  the normalized Ricci scalar with respect to the cosmological constant $\Lambda$. Note that this type of models are shown to be capable of reproducing the whole cosmological history \cite{OdintsovSGS:2017}. Here we have omitted an inflationary term $F_\mathrm{inf} $ as must become negligible at late times, when the observational data that we use in this paper turns out relevant.

For our purposes, we can consider a flat Friedmann-Lema\^itre-Robertson-Walker metric:
\be
ds^2 = -dt^2 + a^2(t)\,\delta_{ij}dx^idx^j\ . \label{FLRW}
\ee
Then, the FLRW equations lead to \cite{OdintsovOik_UniAx:2019,OdintsovOik_Axion:2020}:
 \begin{eqnarray}
 3 H^2F_R&=&\frac{RF_R-F}{2}-3H\dot{F}_R+\kappa^2\left(
\rho+\frac{1}{2}\dot{\phi}^2+V(\phi)\right)\, ,\label{eqn1} \\
-2\dot{H}F_R&=&\kappa^2(\rho+p+\dot{\phi}^2)+\ddot{F}_R-H\dot{F}_R \, ,\label{eqn2}\\
\ddot{\phi}&+&3H\dot{\phi}+V'(\phi)=0, \qquad V(\phi)=\frac12 m_a^2 \phi^2 ,
\label{eqphi}
 \end{eqnarray}
where $F_R=\frac{\partial F}{\partial R}$,  the dot denotes differentiation with respect to the cosmic time $t$, $H=\dot a/a$,
$\rho$ is the matter energy density and $p$ the pressure, which gathers all the species of the universe which satisfy the continuity equation:
\begin{equation}\dot\rho=-3H(\rho+p)
  \label{cont}\end{equation}
Hence, the set of equations given in (\ref{eqn1})-(\ref{eqphi}), together with the continuity equations (\ref{cont}) provide the complete description of the universe evolution. We are gonna consider now two possible scenarios: the first one in absent of the axion field, such that dark matter is described just by a pressureless fluid, which also includes baryons, dark  matter (non-relativistic)  and radiation (relativistic particles):
\begin{equation}
\rho=\rho_m+\rho_r=\rho_b+\rho_{dm}+\rho_r,\qquad p_m=0,\qquad p_r=\frac13\rho_r\ .
 \label{rho}\end{equation}
Whereas in the second scenario, dark matter is described by the axion field $\phi$. \\

Let us start by analysing the first case, where ${\cal L}_{\phi}$ is absent in the action (\ref{Act1}). Then, the evolution of the different species is provided by the continuity equation (\ref{cont}), which
  for dust matter $\rho_m$  and radiation $\rho_r$ yield:
  \begin{equation}
 \rho=\rho_m^0a^{-3}+ \rho_r^0a^{-4}=\rho_m^0(a^{-3}+X_r a^{-4})\, .
 \label{rho2}\end{equation}
 Here $a=1$, $\rho_m^0$ and $\rho_r^0$ are the present values of the scale factor and matter densities, while we have assumed that dark matter is included in $\rho_m=\rho_b+\rho_{dm}$ and the ratio among densities is estimated from Planck data \cite{Planck13,Planck18}:
  \begin{equation}
  X_r=\frac{\rho_r^0}{\rho_m^0}=2.9656\cdot10^{-4}\ . \label{Xrm}
  \end{equation}
In this first scenario, the set of equations (\ref{eqn1}) and (\ref{eqn2}) can be rewritten in a dynamical system form together with the continuity equation (\ref{cont}), leading to \cite{OdintsovSGS:2017,OdintsovSGSFlog:2019}:
 \begin{eqnarray}
 \frac{dH}{d\log a}&=&\frac{R}{6H}-2H, \label{eqH}\\
 \frac{dR}{d\log a}&=&\frac1{F_{RR}}\bigg(\frac{\kappa^2\rho}{3H^2}-F_R+\frac{RF_R-F}{6H^2}\bigg)\ .
 \label{eqR}\end{eqnarray}
Note that the exponential model (\ref{FR2})  turns out the $\Lambda$CDM model in the limit $\beta\to\infty$ and in the limit of high curvature: $R\gg\Lambda/\beta$. Hence,
physical solutions for this $F(R)$ model should tend asymptotically to $\Lambda$CDM
solutions at large redshifts, such that the corresponding viable solutions from the system (\ref{eqH}),
(\ref{eqR}) must accomplish the $\Lambda$CDM-like asymptotic behavior at early times (before and near
the recombination, but much later the inflationary era). In other words, at redshifts
$z\to\infty$ this model should mimic the $\Lambda$CDM model
\cite{OdintsovSGS:2017} that behaves at large $z$ as
\begin{equation}
 \frac{H^2}{(H^{*}_0)^2}=\Omega_m^{*} \big(a^{-3}+ X_r^{*}a^{-4}\big)+\Omega_\Lambda^{*},\qquad
 \frac{R}{2\Lambda}=2+\frac{\Omega_m^{*}}{2\Omega_\Lambda^{*}}a^{-3}, \qquad
 a\to0.
  \label{asymLCDM}\end{equation}
Here the index $*$ refers to the parameters as provided by the
$\Lambda$CDM model. In particular, $\Omega_\Lambda^{*}=\frac\Lambda{3(H^{*}_0)^2}$ and
$H^{*}_0$ is the Hubble constant in the $\Lambda$CDM asymptotic relations (\ref{asymLCDM}). These relations (\ref{asymLCDM}) are used as the initial conditions when integrating
numerically the system of equations  (\ref{eqH})-(\ref{eqR}) over the variable $x=\log a=-\log(z+1)$  from the initial point $x_i$. This point corresponds to an epoch, when the
factor $\varepsilon=e^{-\beta {\cal R}(x_i)}$ was in the interval $(10^{-9},10^{-7})$,
such that from Eq.~(\ref{asymLCDM}), the initial starting point $x_i$ can be expressed as \cite{OdintsovSGS:2017}:
 \begin{equation}
  x_i=\frac13\log\frac{\beta\Omega_m^{*}}{2\Omega_\Lambda^{*}(\log\varepsilon^{-1}-2\beta)}\,.
  \label{xini}\end{equation}
 In this case, the $F(R)$ model mimics $\Lambda$CDM at early
times (before and near $x_i$) but its late-time evolution deviates from $\Lambda$CDM model despite the same initial conditions for both hold. Consequently, the above parameters $H^{*}_0$,
$\Omega_m^{*}$ differ from the corresponding parameters of the $F(R)$ model as measured today:  $ H_0=H(t_0)$, $\Omega_m^0=\kappa^2\rho_m(t_0)/(3 H_0^2)$. Nevertheless, the parameters for both models are connected as far as the energy density measured today and the cosmological constant remain the same for both cases \cite{HuSawicki07,OdintsovSGS:2017}:
 \begin{equation}
 \Omega_m^0H_0^2=\Omega_m^{*}(H^{*}_0)^2=\frac{\kappa^2}3\rho_m(t_0),
 \qquad  \Omega_\Lambda H_0^2=\Omega_\Lambda^{*}(H^{*}_0)^2=\frac{\Lambda}3\ .
  \label{H0Omm}\end{equation}
Hence, for this first scenario  we are analysing the cosmological evolution for the $F(R)$  model (\ref{FR2}) by considering the following free parameters: $\beta$, $H_0$, $\Omega_m^0$, $\Omega_\Lambda$.\\

Let us now introduce the second scenario. We now consider the model (\ref{FR2}) but including the axion $\phi$ instead of dark
matter just as a pressureless fluid. In such a case, we have an additional free parameter $m_a$ in the potential (\ref{eqphi}). For convenience, we use its dimensionless analog:
 \begin{equation}
 \mu_a=\frac{m_a}{H_0}\, .
  \label{mua}\end{equation}
In this scenario the axion oscillates but its energy density $\rho_a=\frac{1}{2}\dot{\phi}^2+V(\phi)$ behaves like
cold matter \cite{OdintsovOik_UniAx:2019,OdintsovOik_Axion:2020}:
\begin{equation}
\rho_a\simeq\rho_a^0a^{-3}\ .
 \label{rhoax}\end{equation}
In this case, equations (\ref{rho}) and (\ref{rho2}) become:
\begin{equation}
\rho=\rho_b+\rho_r=\rho_b^0a^{-3}+ \rho_r^0a^{-4}\,.
 \label{rho3}\end{equation}
 For  the present time baryon density, we will use the Planck 2018 estimation \cite{Planck18} given by:
  $$X_b=\frac{\rho_b^0}{\rho_m^0}\simeq0.1574\,.$$
The system of equations (\ref{eqn1})\,--\,(\ref{eqphi}) for the $F(R)$ model with the axion field (\ref{Act1}) can be also rewritten in a dynamical system form, as done above for the first scenario. To do so, equation (\ref{eqn1}) can be rewritten similarly to Eq.~(\ref{eqR}) (remind that $x=\log a$) as:
\begin{equation}
 \frac{dR}{dx}=\frac1{F_{RR}}\bigg[\frac{\kappa^2}{3H^2}\Big(\rho+\frac{\dot{\phi}^2+ m_a^2 \phi^2}{2}\Big)-F_R+\frac{RF_R-F}{6H^2}\bigg],
 \label{eqRax}
\end{equation}
%Thus, dynamics of the model (\ref{Act1}) is described by the system of equations (\ref{eqphi}), (\ref{eqH}), (\ref{cont}) or (\ref{rho3}) and (\ref{eqRax}).
By redefining the Hubble parameter and the scalar field $\phi$ (and $\dot\phi$) as dimensionless functions:
\begin{equation}
E=\frac{H}{H_0^{*}},\qquad \Phi=\kappa\phi,\qquad \Psi=\frac{\kappa\dot\phi}{H_0^{*}}\, ,
   \label{EPhi}\end{equation}
the equations  (\ref{eqphi}), (\ref{eqH}) and (\ref{eqRax}) are expressed as:
\begin{eqnarray}
\frac{dE}{dx}&=&\Omega_\Lambda^{*}\frac{{\cal R}}{E}-2E\ ,  \label{eqH2}\\
\frac{d{\cal R}}{dx}&=&\frac{e^{\beta{\cal
R}}}{\beta^2}\bigg[\frac{\Omega_m^{*}(X_ba^{-3}+ X_ra^{-4})+\frac16(\Psi^2+\mu_a^2\Phi^2)} {E^2}-1+\beta e^{-\beta {\cal
R}}+\Omega_\Lambda^{*}\frac{1-(1+\beta {\cal R})\,e^{-\beta {\cal R}}}{E^2}\bigg]\ .
  \label{eqRax2}\\
\frac{d\Phi}{dx}&=&\frac1{E}\Psi,\qquad \frac{d\Psi}{dx}=-3\Psi-\frac{\mu_a^2}{E}\Phi\,.
 \label{eqphi2}
  \end{eqnarray}
As above, the exponential model (\ref{FR2}) with the axion field should also tend asymptotically to the $\Lambda$CDM model at large redshifts. Then, by assuming the early evolution (\ref{rhoax}) for the axion energy density, we can also use the above asymptotic initial conditions (\ref{asymLCDM}) at the initial point $x_i$ (\ref{xini}) and integrate numerically the system (\ref{eqH2})-(\ref{eqphi2}) in the interval $x_i\le x\le0$. Note that the initial conditions at $x=x_i$ for $E$ and ${\cal R}$ are provided by the Eqs.~(\ref{asymLCDM}). Nevertheless, the initial conditions for the axion field require a clarification. It is natural to choose $\phi(x_i)$ and $\dot\phi(x_i)$ in such a way that a smooth transition from a
cold matter evolution (\ref{rhoax}) of the axion density before $x_i$ to its real evolution
$\rho_a(x)$ after $x_i$, is followed. These conditions of smoothness take the form:
 \begin{equation}
\rho_a(x_i)=\rho_{dm}^0a^{-3}\big|_{x_i}=\rho_{dm}^0e^{-3x_i},\qquad \frac{d\rho_a}{dx}\Big|_{x_i}=-3\rho_{dm}^0e^{-3x_i}\, .
   \label{smooth}\end{equation}

\begin{figure}[bh]
  \centerline{\includegraphics[scale=0.71,trim=4mm 4mm 5mm 5mm]{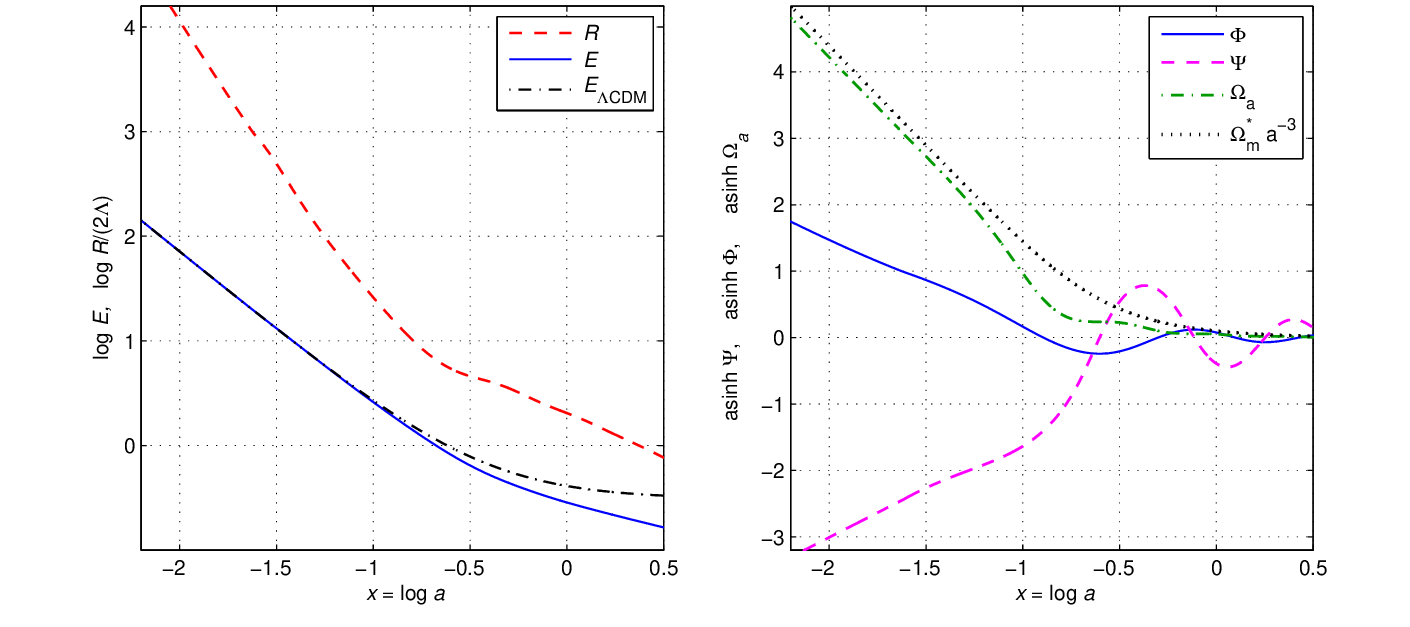}}
\caption{Evolution of $E$, ${\cal R}$ (left) and axion amplitudes $\Phi$, $\Psi$,
$\Omega_a(x)$ (right) for the exponential $F(R)$ model (\ref{FR2}) with axion. The values for the model
parameters are given in Table~\ref{Estim}. }
  \label{F1}
\end{figure}

In terms of the variables defined in (\ref{EPhi}), the first condition in (\ref{smooth}) is translated to:
 $$ (\Psi^2+\mu_a^2\Phi^2)\big|_{x_i}=6\Omega_m^*(1-X_b)\,e^{-3x_i}\,,
 $$
whereas  the second condition is obtained by Eqs.~(\ref{eqphi2}), leading to:
$$\frac{d}{dx}(\Psi^2+\mu_a^2\Phi^2)\big|_{x_i}=-6\Psi^2\big|_{x_i}\ .$$
Hence, from the smoothness relations (\ref{smooth}), the following initial conditions for $\Phi$ and $\Psi$ are finally obtained:
 \begin{equation}
\Psi^2\big|_{x_i}=\mu_a^2\Phi^2\big|_{x_i}=3\Omega_m^*(1-X_b)\,e^{-3x_i}\, .
   \label{Psiini}\end{equation}
Fig.~\ref{F1} depicts the results of integrating the system (\ref{eqH2})-(\ref{eqphi2}) once the above initial conditions are assumed. In the left
panel, the logarithms of the normalized Ricci scalar ${\cal R}(x)$ and the Hubble parameter
$E(x)$ are shown, including also the Hubble parameter for the $\Lambda$CDM model (\ref{asymLCDM}). Both evolutions coincide close to $x<x_i$ but deviate at current times. The corresponding model
free parameters are those shown in Table~\ref{Estim} as the best fit values for this scenario. In the right panel of Fig.~\ref{F1} the axion amplitudes (\ref{EPhi}) $\Phi(x)$, $\Psi(x)$
are shown with the normalized axion energy density $\Omega_a(x)=\frac{\kappa^2}{3H_0^*{^2}}\rho_a(x)$ compared with $\Omega_m^* a^{-3}$. The functions  $\Phi(x)$, $\Psi(x)$ oscillate and grow rapidly as $a\to 0$, so the inverse hyperbolic functions are used here  (asinh$\,x=\log(x+\sqrt{1+x^2})$.

\smallskip

In the next sections, we confront the above models with the observational data and compare them with the $\Lambda$CDM model.

%%%%%%%%%%%%%%%%%%%%%%%%%%%%%%%%%%%%%%%%%%%%%%%%%%%%%%%%%%%%
\section{Observational data}\label{Observ}
%%%%%%%%%%%%%%%%%%%%
In this section, the corresponding observational data for testing the above scenarios for the $F(R)$ model (\ref{FR2}), namely  (i)  the usual dark matter as a pressureless fluid and (ii)  the axion field $\phi$, are shown up in order to estimate the viability of both models and obtain the best fit for the free parameters. The comparison with observational data includes the following sources: (a)
Pantheon sample of Type Ia supernovae (SNe Ia) data \cite{Scolnic17}; (b) measurements
of the Hubble parameter $H(z)$ from Cosmic Chronometers (CC), (c) Cosmic
Microwave Background radiation (CMB) data and (d) Baryon Acoustic Oscillations (BAO). A detailed description of the corresponding
data analysis and methods followed can be found in Refs.~\cite{OdintsovSGSFlog:2019,OdintsovSGS:2022,OdintsovOS:2023}.\\

The tests are performed by evaluating the $\chi^2$ function, which is given by:
 \begin{equation}
  \chi^2\equiv\chi^2_\mathrm{tot}=\chi^2_\mathrm{SN}+\chi^2_H+\chi^2_\mathrm{CMB}+\chi^2_\mathrm{BAO}\,.
 \label{chitot} \end{equation}
and includes the contributions from SNe Ia, CC $H(z)$ data, CMB  and BAO comparisons. The  SNe Ia  $\chi^2$ function is given by:
\begin{equation}
\chi^2_{\mathrm{SN}}(\theta_1,\dots)=\min\limits_{H_0} \sum_{i,j=1}^{N_\mathrm{SN}}
 \Delta\mu_i\big(C_{\mathrm{SN}}^{-1}\big)_{ij} \Delta\mu_j,\qquad
 \Delta\mu_i=\mu^\mathrm{th}(z_i,\theta_1,\dots)-\mu^\mathrm{obs}_i\ ,
 \label{chiSN}\end{equation}
 where  $N_{\mathrm{SN}}=1048$ datapoints of the distance moduli $\mu_i^\mathrm{obs}$ at redshifts
$z_i$ as provided by the Pantheon sample database \cite{Scolnic17}, while  $\theta_j$ are free
model parameters, $C_{\mbox{\scriptsize SN}}$ is the covariance matrix  \cite{Scolnic17}
and $\mu^\mathrm{th}$ are the theoretical values, which are calculated as follows:
\begin{equation}
 \mu^\mathrm{th}(z) = 5 \log_{10} \frac{(1+z)\,D_M(z)}{10\mbox{pc}},\qquad D_M(z)= c \int\limits_0^z\frac{d\tilde z}{H(\tilde
 z)}.    \label{muDM}
\end{equation}
For evaluating the $\chi^2_{\mathrm{SN}}$ function (\ref{chiSN}), the Hubble constant $H_0$ (or equivalently the
``asymptotical'' constant $H_0^*$) is considered  as a
nuisance parameter.
For the Hubble parameter data  $H(z)$ we use here $N_H=32$ datapoints of Cosmic Chronometers (CC) given in Refs.~\cite{HzData}, i.e. measured as $H (z)= \frac{\dot{a}}{a} \simeq -\frac{1}{1+z} \frac{\Delta z}{\Delta t}$ from different ages $\Delta t$ of galaxies with close redshifts $\Delta z$. The corresponding $\chi^2$ function for CC $H(z)$ data yields:
\begin{equation}
    \chi_H^2(\theta_1,\dots)=\sum_{j=1}^{N_H}\bigg[\frac{H(z_j,\theta_1,\dots)-H^{obs}(z_j)}{\sigma _j}  \bigg]^2.
    \label{chiH}
\end{equation}
From the CMB  we use observational parameters obtained from Planck 2018 data
\cite{Planck18} in the form \cite{ChenHuangW2018}:
  \begin{equation}
  \mathbf{x}=\big(R,\ell_A,\omega_b\big),\qquad R=\sqrt{\Omega_m^0}\frac{H_0D_M(z_*)}c,\quad
 \ell_A=\frac{\pi D_M(z_*)}{r_s(z_*)},\quad\omega_b=\Omega_b^0h^2\ .
 \label{CMB} \end{equation}
Here $z_*$ is the photon-decoupling redshift, $D_M$ is the comoving distance
(\ref{muDM}), $h=H_0/[100\,\mbox{km}\mbox{s}^{-1}\mbox{Mpc}^{-1}]$,  $r_s(z)$ is the
comoving sound horizon. Details on the way to obtain $r_s(z)$ and other parameters are given in Appendix. The corresponding $\chi^2$ function for the CMB data is given by:
 \begin{equation}
\chi^2_{\mbox{\scriptsize CMB}}=\min_{\omega_b}\Delta\mathbf{x}\cdot
C_{\mathrm{CMB}}^{-1}\big(\Delta\mathbf{x}\big)^{T},\qquad \Delta
\mathbf{x}=\mathbf{x}-\mathbf{x}^{Pl}
 \label{chiCMB} \end{equation}
where the estimations given in Ref.~\cite{ChenHuangW2018} are used:
 $\mathbf{x}^{Pl}=\big(R^{Pl},\ell_A^{Pl},\omega_b^{Pl}\big)=\big(1.7428\pm0.0053,\;301.406\pm0.090,\;0.02259\pm0.00017\big)
$, obtained from Planck 2018 data \cite{Planck18} with free amplitude for
the lensing power spectrum. The covariance matrix $C_{\mathrm{CMB}}=\|\tilde
C_{ij}\sigma_i\sigma_j\|$ is described in Ref.~\cite{ChenHuangW2018}.\\

For the baryon acoustic oscillations (BAO) data the following two magnitudes are considered:
\begin{equation}
d_z(z)= \frac{r_s(z_d)}{D_V(z)}\, ,\qquad A(z) = \frac{H_0\sqrt{\Omega_m^0}}{cz}D_V(z)\,
, \label{dzAz}
\end{equation}
where $D_V(z)=\big[{cz D_M^2(z)}/{H(z)} \big]^{1/3}$, $z_d$ being the redshift at the
end of the baryon drag era. Here we use  21 BAO data points for $d_z(z)$ and 7 data
points for $A(z)$ (as given in Table~\ref{TBAO} in  Appendix) in the following $\chi^2$
function:
\begin{equation}
\chi^2_{\mathrm{BAO}}(\Omega_m^0,\theta_1,\dots)=\Delta d\cdot C_d^{-1}(\Delta d)^T +
\Delta { A}\cdot C_A^{-1}(\Delta { A})^T\, . \label{chiBAO}
\end{equation}
Here, $\Delta d_i=d_z^\mathrm{obs}(z_i)-d_z^\mathrm{th}(z_i,\dots)$, $\Delta
A_i=A^\mathrm{obs}(z_i)-A^\mathrm{th}(z_i,\dots)$, $C_{d}$ and $C_{A}$ are the
covariance matrices for the correlated BAO data \cite{Percival:2009,Blake:2011}.

%%%%%%%%%%%%%%%%%%%%%%%%%%%%%%%%%%%%%%%%%%%%%%%%%%%%%%%
\section{Results and discussion}\label{Results}
%%%%%%%%%%%%%%%%%%%%%%%%%%%%%%%%%%%%%%%%%%%%%%%%%%%%%%

Let us now fit the  the exponential $F(R)$ model (\ref{FR2}) with the
observational data described in the previous section. To do so, we minimize the $\chi^2$ function
(\ref{chitot}), including SNe Ia data  (\ref{chiSN}), CC $H(z)$ data (\ref{chiH}), CMB
(\ref{chiCMB}) and BAO (\ref{chiBAO}) contributions. This function is calculated in the
space of free model parameters  with flat priors within their natural limitations
(positive values for $\beta$, $\Omega_m^0$, $\Omega_\Lambda$, $H_0$).

\begin{figure}[th]
  \centerline{\includegraphics[scale=0.71,trim=4mm 4mm 5mm 5mm]{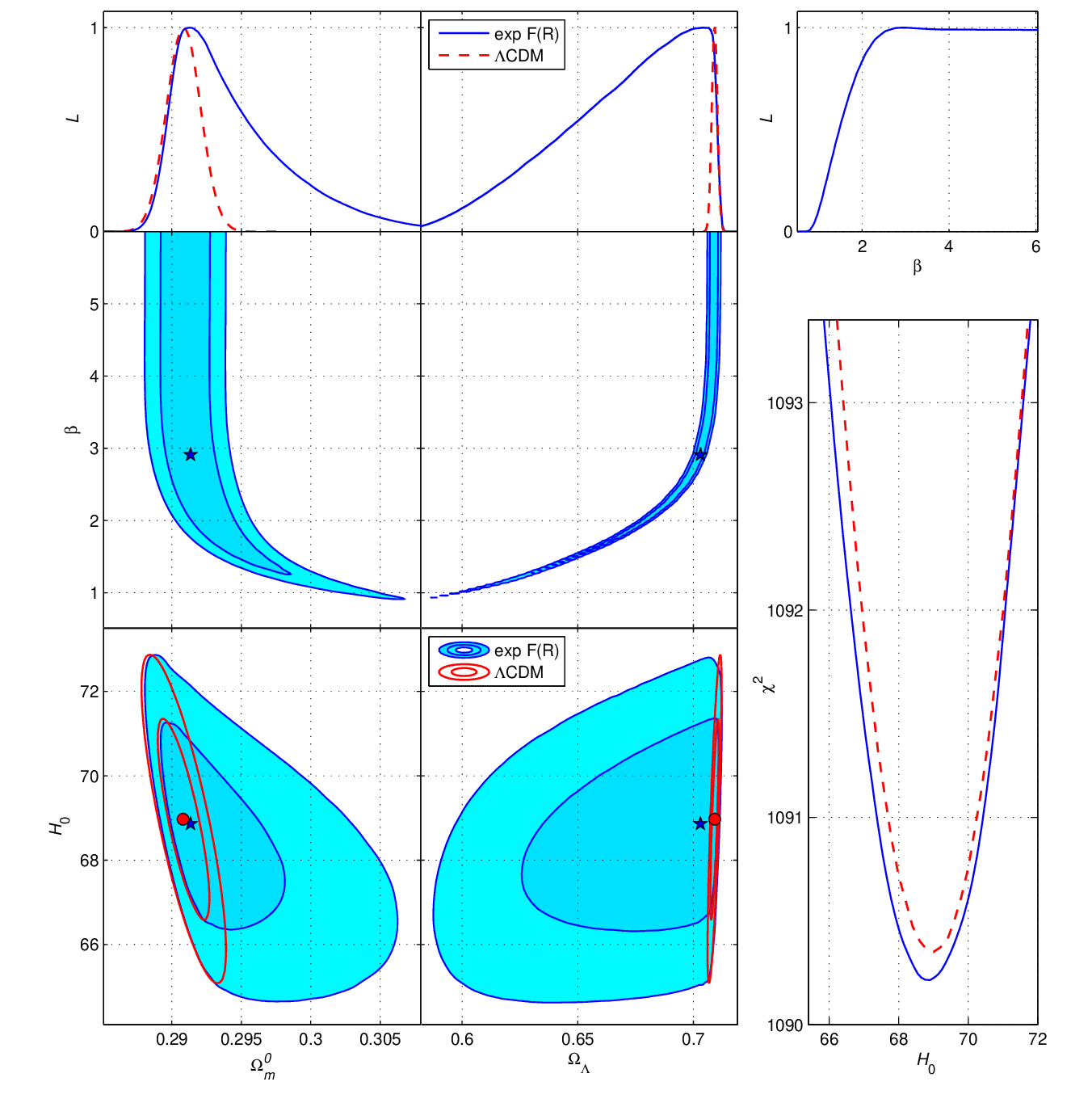}}
\caption{Contour plots for the $1\sigma$, $2\sigma$ CL (bottom left panels) as well as the likelihood functions $ {\cal L}(\theta_i)$ (in the top panels) for the  exponential
$F(R)$ model without axion in comparison with the $\Lambda$CDM model in the
$\Omega_m^0-H_0$ plane and other panels. The stars for the $F(R)$ model and  circles for
the $\Lambda$CDM denote the minimum points of the $\chi^2$ functions. The one-parameter
distribution $\chi^2(H_0)$ is depicted in the bottom-right panel for both cases.}
  \label{F2}
\end{figure}

In the first scenario (in absence of the axion field) dark matter is included as cold matter evolving as (\ref{rho2}):
$
\rho_m=\rho_b+\rho_{dm}=\rho_m^0a^{-3}.
$
For this scenario, the system of equations (\ref{eqH})-(\ref{eqR}) is integrated and then the corresponding $\chi^2$
function (\ref{chitot}) is obtained by using Monte Carlo Markov Chains (MCMC). The
 contour plots in panels of 2 parameters and likelihoods for $\beta$,
$\Omega_m^0$, $\Omega_\Lambda$, $H_0$ are depicted in Fig.~\ref{F2}.

%We use here the real values $H_0=H(t_0)$, $\Omega_m^0=\frac{\kappa^2}{3H_0^2}\rho_m(t_0)$  and  $\Omega_\Lambda$ keeping in mind that $H(z)=H_0^*E(z)$ with the true value for the Hubble parameter today being $H_0=H_0^*E(z=0)$ and also the relation (\ref{H0Omm}) for the matter density.

The two-parameter distributions $\chi^2_\mathrm{tot}(\theta_i,\theta_j)$ correspond to $1\sigma$ (68.27\%) and $2\sigma$ (95.45\%)
confidence levels (CL). These distributions are marginalised over all the remaining free
parameters. For instance, in the bottom-left panel of Fig.~\ref{F2} the corresponding $1\sigma$  and $2\sigma$ CL regions are shown for
$$\chi^2(\Omega_m^0,H_0)=\min\limits_{\beta,\Omega_\Lambda}\chi^2.$$
 In other panels of Fig.~\ref{F2} a similar approach is followed, the functions $\chi^2(\theta_i,\theta_j)$ reach their absolute minima at the points
marked by stars (or circles for the $\Lambda$CDM model).

Here one-parameter distributions are also obtained by minimising over all the remaining
model parameters. For $\Omega_m^0$,  $\Omega_\Lambda$ and $\beta$  the likelihood
functions are depicted in Fig.~\ref{F2}. They are related with the corresponding
one-parameter distributions as:
 $${\cal L}(\Omega_m^0)\sim\exp(-\chi^2_\mathrm{tot}(\Omega_m^0)/2).$$

In the bottom-right panel of  Fig.~\ref{F2}, the one-parameter distribution
$\chi^2(H_0)$ is shown in comparison with the one from the  $\Lambda$CDM model (\ref{asymLCDM}).
%For the $\Lambda$CDM model we use the Hubble parameter
%\begin{equation}
%H^2=H_0^2\Big[\Omega_m^0 \big(a^{-3}+ X_ra^{-4}\big)+\Omega_\Lambda\Big]
  %\label{LCDM}\end{equation}
  %and $\Omega_\Lambda=1-\Omega_m^0(1+ X_r)$ in the considered spatially flat case.

As shown in Fig.~\ref{F2}, the best fit values $\min\chi^2\simeq1090.21$ for the
$F(R)$ model is slightly better in comparison to the $\Lambda$CDM model, which returns a result given by $1090.35$.
This tiny difference lies on the fact that the $\chi^2$ for the $F(R)$ model reaches its minimum at
 $\beta\simeq2.94$, whereas the $\Lambda$CDM is recovered at the limit $\beta\to\infty$.

The best fit $\Omega_m^0=0.2913^{+0.0035}_{-0.0015}$ for the $F(R)$ model is close to the $\Lambda$CDM estimation, but the $1\sigma$ error is larger. A similar (and larger)  difference may be seen for
$\Omega_\Lambda$ in Fig.~\ref{F2}. This is connected with the additional degrees of freedom within the exponential $F(R)$ model for finite $\beta$ in comparison to the $\Lambda$CDM model. However, the estimations of the Hubble constant $H_0$ for these two models are rather close.

%Remind that we use the ``true'' parameters $\Omega_m^0$,   $\Omega_\Lambda$ and $H_0=H(t_0)$ in Fig.~\ref{F2}. These values  should be differed from the ``asymptotic'' parameters $\Omega_m^*$, $H_0^*$, $\Omega_\Lambda^*$ (\ref{asymLCDM}) keeping in mind their connection (\ref{H0Omm}).

The best fit, the $1\sigma$ error for the other parameters and the values $\min\chi^2$ are
gathered in Table \ref{Estim}. They are determined by the one-parameter distributions or likelihoods ${\cal L}(\theta_j)$.

%%{\small
\begin{table}[ht]
%\begin{center}
\begin{tabular}{|l|c|c|c|c|c|c|c|}
\hline  Model &   $\min\chi^2/d.o.f$& AIC & $\Omega_m^0$& $\Omega_\Lambda$& $H_0$ &  $\beta$ & $\mu_a$   \\
\hline
Exp $F(R)$ & 1090.21 /1108 & 1098.21& $0.2913^{+0.0035}_{-0.0015}$ & $0.703^{+0.007}_{-0.047}$ & $68.84^{+1.75}_{-1.64}$& $2.94^{+\infty}_{-1.325}$ & -  \rule{0pt}{1.1em}  \\
\hline
Exp $F(R)$ + axion & 1089.53 /1107 & 1099.53& $0.2967^{+0.0062}_{-0.0061}$ & $1.095^{+0.076}_{-0.066 }$ & $68.78^{+1.68}_{-1.69}$& $0.785^{+0.123}_{-0.084}$ & $8.72^{+0.89}_{-0.91}$ \rule{0pt}{1.1em}  \\
\hline
$\Lambda$CDM& 1090.35 /1110 & 1094.35& $0.2908^{+0.0013}_{-0.0012}$&$0.7092^{+0.0012}_{-0.0013}$& $68.98^{+1.58}_{-1.60}$& - & -  \rule{0pt}{1.1em}  \\
\hline
 \end{tabular}
 \caption{Best  fits  for the free parameters and $\min\chi^2$ for the two dark matter scenarios within the exponential $F(R)$ model
(\ref{Act1}) in comparison with the $\Lambda$CDM model.}
%\end{center}
\label{Estim}
\end{table}

In Table \ref{Estim} one can note that the $F(R)$ model without axion has slightly
better results with respect to the $\Lambda$CDM model regarding the $\min\chi^2$. While
for the model with the axion this difference is smaller. However, $\Lambda$CDM model
still presents a better goodness of the fits when taking into account the number of the
degrees of freedom for each model, provided that $N_p=4$ and $N_p=5$ for the $F(R)$ models without and with the axion respectively, while $N_p=2$ for  the$\Lambda$CDM model.
Then, it is very convenient for comparing all the models to consider the Akaike information criterion
\cite{Akaike}:
 \begin{equation}
 \mbox{AIC} = \min\chi^2_{tot} +2N_p.
  \label{AIC}\end{equation}
Hence, the more free parameters of the model, the better fits yield, but at the price of
increasing the degrees of freedom of the model and consequently leading to a larger
Akaike parameter.\\

The second scenario of the exponential $F(R)$ model (\ref{Act1}) with the axion field
yields the lowest $\min\chi^2$ among the models, as shown in Table \ref{Estim}.
However, this scenario involves an additional parameter $ \mu_a=m_a/H_0$ (\ref{mua}),
that leads consequently to a larger value for the AIC (\ref{AIC}).

\begin{figure}[th]
  \centerline{\includegraphics[scale=0.71,trim=4mm 4mm 5mm 5mm]{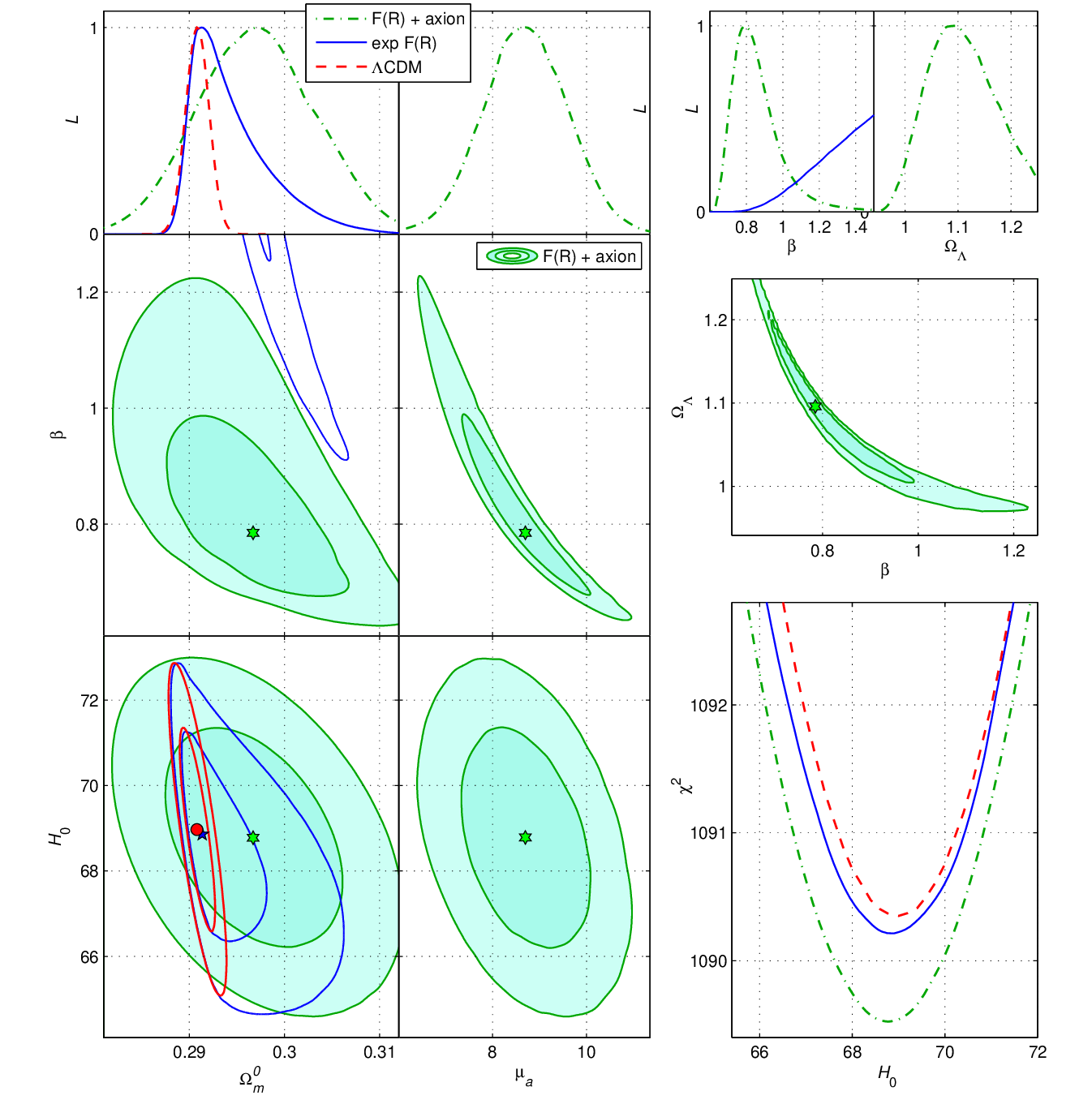}}
\caption{ Results for the exponential $F(R)$ model (\ref{Act1}) with the axion field: contour plots of the projection of the
$\chi^2$ function (marginalised over the other parameters) show the $1\sigma$ and $2\sigma$ CL (bottom left panels), likelihoods for each parameter (top panels) and $\chi^2(H_0)$ (bottom right panel). All the plots also include the case without axion and the $\Lambda$CDM model, following the same notation as Fig.~2. The minimum points for the $\chi^2$ functions of two parameters are marked with green stars (axion model), blue star (without axion field) and circles ($\Lambda$CDM model).}
  \label{F3}
\end{figure}

The $F(R)$ model (\ref{Act1}) with the axion is described by the system of equations
(\ref{eqH2})\,--\,(\ref{eqphi2}), where the initial conditions
(\ref{asymLCDM})-(\ref{Psiini}) are assumed. All the results of the calculations for this scenario are shown
in Fig.~\ref{F3}, where same notation as in Fig.~\ref{F2} is followed.

In particular,  in the bottom-left panel of Fig.~\ref{F3} the $1\sigma$ and
$2\sigma$ CL are depicted for the axion model together with the same contours plots for the other two models
(shown above in Fig.~\ref{F2}) in the $\Omega_m^0-H_0$ plane, which are two parameters in
common for all the cases. We see that for the  $F(R)$ model with the axion the
$1\sigma$ and $2\sigma$ CL domains are wider and slightly shifted along the
$\Omega_m^0$ axis, whereas the observed $H_0$ dependence is very close for 3 models. In the bottom-right panel in Fig.~\ref{F3},  the one-parameter
distributions $\chi^2(H_0)$ show similar shapes but differ in the minimum values of
$\chi^2$: the lowest $\min\chi^2$ is for the axion $F(R)$ model.

The most important difference between the two considered exponential $F(R)$ models lies on the fits for the parameter $\beta$ (see the $\Omega_m^0-\beta$ plane in
Fig~\ref{F3}). While the model without the axion works successfully for large $\beta$, where the best
fit is given by $\beta=2.94^{+\infty}_{-1.325}$), recovering $\Lambda$CDM model at $\beta\rightarrow\infty$, the presence of the axion provides a finite  $1\sigma$ (and $2\sigma$ ) CL, where $\beta=0.785^{+0.123}_{-0.084}$, moving away from the $\Lambda$CDM model behaviour. As far as $\beta$ is small enough, $\Omega_\Lambda$ behaves as a free model parameter, as one can see in Fig~\ref{F3} and in Table \ref{Estim}, where its value for the axion model is rather large. This deviation from the $\Lambda$CDM scenario leads to certain differences in the predictions of the observable parameters, in particular, $\Omega_m^0$ is also slightly enlarged in the axion model, but within the $1\sigma$ region. Remind that $\beta$ in the exponential $F(R)$ models is not a direct observable parameter. In addition, the parameter $\mu_a$ associated to the axion field has the best fit value
$\mu_a=8.72^{+0.89}_{-0.91}$. We may conclude that the $F(R)$ model (\ref{Act1}) with the axion field successfully describe the observational data as far as  $\mu_a$,  $\beta$ and other free model parameters lie inside the limited error boxes (see  Table \ref{Estim}).

%%%%%%%%%%%%%%%%%%%%%%%%%%%%%%%%%%%%%%
\section{Conclusions}
\label{conclusions}
%%%%%%%%%%%%%%%%%%%%%%%%%%%%%%%%%%%%

In this paper exponential $F(R)$ has been considered together with two different descriptions for dark matter. In the first scenario, an effective pressureless fluid is assumed to describe the dark matter component, as usually followed in most of cosmological analysis. For the second case, an axion field is considered to play the role of dark matter. The latter has been widely analysed previously, since axion fields arise naturally in quantum field theories, specially for solving the CP problem of QCD. In addition, axions may play the role of dark matter when settle down to the minimum of its potential, turning out a good candidate to be searched in the laboratory. Hence, the aim of this paper has been to compare both cases to find out which one might provide better fits when tested with observational data and at the same time, test the viability of exponential gravity in comparison to the $\Lambda$CDM model\\

Then, both scenarios have been tested with different sources of observational data, showing similar good fits, also in comparison to the $\Lambda$CDM model. Nevertheless, one can note that while in the case of standard dark matter, the 1-$\sigma$ region contains in fact the $\Lambda$CDM model, this is not the case for the axion dark matter field, where the 1-$\sigma$ region for the $\beta$ is finite (remind that $\Lambda$CDM is recovered in the limit $\beta\rightarrow\infty$). While this is not statistically significant, it may point to deviations from GR in case that the nature of dark matter relies on an axion field. Nevertheless, the goodness of the fits, based on the Akaike information criterion that counts the number of free parameters of a particular model, shows a slightly better value for the $\Lambda$CDM model in comparison to the others. Nevertheless, such a small difference on the Akaike parameter is not statistically significant either.\\

 In addition, the axion model might provide some way to alleviate the Hubble tension problem, since its deviation from a standard dust fluid might contribute to modify the cosmological evolution before/during the recombination epoch, similarly to Early Dark Energy models, which are known to provide some relief to the Hubble tension problem. Moreover, other open windows to check the possible existence of axion fields as dark matter candidates might lie on the effects that may be induced on the black holes' shadows when considering them surrounding by this type of dark matter, a very active and current field nowadays \cite{Vagnozzi:2022moj}.\\

Hence, one might conclude that axions fields together with some viable models of $F(R)$ gravity provide a reliable description of the universe evolution, including inflation and late-time acceleration. As shown by the statistics, the model fits well the observational data, providing similar fits as the $\Lambda$CDM model. Moreover, as shown by the results of the fits, the discovery of the axion in the laboratory might point to modifications of the Hilbert-Einstein action. While such modifications are not of course a definite theory of gravity, they might point the way GR should be corrected and perhaps the way to find a more general and realistic description of both gravitation as the nature of dark matter.

%%%%%%%%%%%%%%%%%%%%%%%%%%%%%%%%%%%%%%
\section*{Appendix}
\label{App}
%%%%%%%%%%%%%%%%%%%%%%%%%%%%%%%%%%%%

To describe CMB an BAO observational data we calculate the comoving sound horizon
$r_s(z)$ as follows \cite{OdintsovSGS:2022,OdintsovOS:2023}:
  \begin{equation}
r_s(z)=  \int_z^{\infty} \frac{c_s(\tilde z)}{H (\tilde z)}\,d\tilde
z=\frac1{\sqrt{3}}\int_0^{1/(1+z)}\frac{da}
 {a^2H(a)\sqrt{1+\big[3\Omega_b^0/(4\Omega_\gamma^0)\big]a}}\ .
  \label{rs2}\end{equation}
We estimate the ratio of baryons and photons $\Omega_b^0/\Omega_\gamma$ using the
relation (\ref{Xrm}) $\rho_\nu=N_\mathrm{eff}(7/8)(4/11)^{4/3}\rho_\gamma$
with $N_\mathrm{eff} = 3.046$, as given by Planck 2018 data \cite{Planck18}.
 We use the estimation of $z_*$  given in Refs.~\cite{ChenHuangW2018,HuSugiyama95}.
The current baryon fraction $\Omega_b^0$ here is considered as the nuisance parameter in
the corresponding $\chi^2_{\mathrm{CMB}}$ function (\ref{chiCMB}).

For BAO data we use 21 BAO datapoints for the magnitude $d_z(z)$ in (\ref{dzAz}) and 7
data points for $A(z)$  given in Table~\ref{TBAO} from
Refs.~\cite{Percival:2009,Kazin:2009,Beutler:2011,Blake:2011,Chuang:2013,Anderson:2013,Ross:2014,Beutler:2016,Chuang:2017,Bourboux:2017,
Zhu:2018,Blomqvist:2019,Hou:2020,Tamone:2020}.
 This table contains some new data points with respect to BAO data from
 Refs.~\cite{OdintsovSGS:2017,OdintsovSGSFlog:2019} and we
excluded from Table~\ref{TBAO} estimates of $d_z$, extracted from repeating or
overlapping galaxy catalogues.
\begin{table}[th]
\centering
 {\begin{tabular}{||l|l|l|l|l|l|l||}
\hline
 $z$  & $d_z(z)$ &$\sigma_d$    & $ A(z)$ & $\sigma_A$   & Survey & Refs.\\ \hline
 0.106& 0.336  & 0.015 & 0.526& 0.028&  6dFGS & \cite{Beutler:2011}\\ \hline
 0.15 & 0.2237 & 0.0084& -    & -    &  SDSS DR7&\cite{Ross:2014} \\ \hline
 0.20 & 0.1905 & 0.0061& 0.488& 0.016&  SDSS DR7&\cite{Percival:2009} \\ \hline
 0.278& 0.1394 & 0.0049& -    & -    &  SDSS LRG&\cite{Kazin:2009} \\ \hline
 0.314& 0.1239 & 0.0033& -    & -    & SDSS LRG &\cite{Blake:2011} \\ \hline
 0.32 & 0.1181 & 0.0026& -    & -    &   DR10,11&\cite{Anderson:2013} \\ \hline
 0.32 & 0.1165 & 0.0024& -    & -    & BOSS DR12&\cite{Chuang:2017} \\ \hline
 0.35 & 0.1097 & 0.0036& 0.484& 0.016& SDSS DR7 &\cite{Percival:2009} \\ \hline
 0.38 & 0.1011 & 0.0011& -    & -    & BOSS DR12&\cite{Beutler:2016} \\ \hline
 0.44 & 0.0916 & 0.0071& 0.474& 0.034&  WiggleZ &\cite{Blake:2011}\\ \hline
 0.57 & 0.0739 & 0.0043& 0.436& 0.017&  BOSS DR9&\cite{Chuang:2013}\\ \hline
 0.57 & 0.0726 & 0.0014& -    & -    &  DR10,11 &\cite{Anderson:2013}\\ \hline
 0.59 & 0.0701 & 0.0008& -    & -    & BOSS DR12&\cite{Chuang:2017} \\ \hline
 0.60 & 0.0726 & 0.0034& 0.442& 0.020&  WiggleZ &\cite{Blake:2011}\\ \hline
 0.61 & 0.0696 & 0.0007& -    & -    & BOSS DR12&\cite{Beutler:2016} \\ \hline
 0.73 & 0.0592 & 0.0032& 0.424& 0.021& WiggleZ  &\cite{Blake:2011} \\ \hline
 0.85 & 0.0538 & 0.0041& -& - &  DR16 ELG       &\cite{Tamone:2020}\\ \hline
 1.48 & 0.0380 & 0.0013& -& - & eBOSS DR16      &\cite{Hou:2020}\\ \hline
 2.0  & 0.0339 & 0.0025& -& - & eBOSS DR14      &\cite{Zhu:2018}\\ \hline
 2.35 & 0.0327 & 0.0016& -& - & DR14 Ly$\alpha$ &\cite{Blomqvist:2019}\\ \hline
 2.4  & 0.0331 & 0.0016& -& - & DR12 Ly$\alpha$ &\cite{Bourboux:2017}\\  \hline
 \end{tabular}
 \caption{BAO data $d_z(z)=r_s(z_d)/D_V(z)$ and $A(z)$ (\ref{dzAz}).}
 \label{TBAO}} \end{table}

\section*{Acknowledgements}

This work was partially supported by projects Ref.~PID2019-104397GB-I00 (SDO) and Ref.~PID2020-117301GA-I00 (DS-CG) funded by MCIN/AEI/10.13039/501100011033 (``ERDF A way of making Europe" and ``PGC Generaci\'on de Conocimiento", Spain) and also by the program Unidad de Excelencia Maria de Maeztu CEX2020-001058-M, Spain (SDO).


\begin{thebibliography}{99}


%Reviews on modified gravity

\bibitem{Nojiri:2010wj}
S.~Nojiri and S.~D.~Odintsov,
%``Unified cosmic history in modified gravity: from F(R) theory to Lorentz non-invariant models,''
Phys. Rept. \textbf{505}, 59-144 (2011) doi:10.1016/j.physrep.2011.04.001
[arXiv:1011.0544 [gr-qc]];
%\bibitem{Nojiri:2017ncd}
S.~Nojiri, S.~D.~Odintsov and V.~K.~Oikonomou,
%``Modified Gravity Theories on a Nutshell: Inflation, Bounce and Late-time Evolution,''
Phys. Rept. \textbf{692}, 1-104 (2017) doi:10.1016/j.physrep.2017.06.001
[arXiv:1705.11098 [gr-qc]].
%\bibitem{Olmo:2011uz}
G.~J.~Olmo,
%``Palatini Approach to Modified Gravity: f(R) Theories and Beyond,''
Int. J. Mod. Phys. D \textbf{20}, 413-462 (2011) doi:10.1142/S0218271811018925
[arXiv:1101.3864 [gr-qc]].
%364 citations counted in INSPIRE as of 10 Sep 2020
%\bibitem{Capozziello:2011et}
S.~Capozziello and M.~De Laurentis,
%``Extended Theories of Gravity,''
Phys. Rept. \textbf{509}, 167-321 (2011) doi:10.1016/j.physrep.2011.09.003
[arXiv:1108.6266 [gr-qc]];
%\bibitem{Capozziello:2015lza}
S.~Capozziello, T.~Harko, T.~S.~Koivisto, F.~S.~N.~Lobo and G.~J.~Olmo,
%``Hybrid metric-Palatini gravity,''
Universe \textbf{1}, no.2, 199-238 (2015) doi:10.3390/universe1020199 [arXiv:1508.04641
[gr-qc]].
%%%
%\bibitem{Clifton:2011jh}
T.~Clifton, P.~G.~Ferreira, A.~Padilla and C.~Skordis,
%``Modified Gravity and Cosmology,''
Phys. Rept. \textbf{513}, 1-189 (2012) doi:10.1016/j.physrep.2012.01.001
[arXiv:1106.2476 [astro-ph.CO]];
%\bibitem{delaCruzDombriz:2012xy}
A.~de la Cruz-Dombriz and D.~Saez-Gomez,
%``Black holes, cosmological solutions, future singularities, and their thermodynamical properties in modified gravity theories,''
Entropy \textbf{14}, 1717-1770 (2012) doi:10.3390/e14091717 [arXiv:1207.2663 [gr-qc]].

%State of the art cosmology

\bibitem{DiValentino:2020vhf}
E.~Di Valentino, L.~A.~Anchordoqui, O.~Akarsu, Y.~Ali-Haimoud, L.~Amendola, N.~Arendse, M.~Asgari, M.~Ballardini, S.~Basilakos and E.~Battistelli, \textit{et al.}
%``Snowmass2021 - Letter of interest cosmology intertwined I: Perspectives for the next decade,''
Astropart. Phys. \textbf{131}, 102606 (2021)
doi:10.1016/j.astropartphys.2021.102606
[arXiv:2008.11283 [astro-ph.CO]];
%\bibitem{DiValentino:2020zio}
%E.~Di Valentino, L.~A.~Anchordoqui, O.~Akarsu, Y.~Ali-Haimoud, L.~Amendola, N.~Arendse, M.~Asgari, M.~Ballardini, S.~Basilakos and E.~Battistelli, \textit{et al.}
%``Snowmass2021 - Letter of interest cosmology intertwined II: The hubble constant tension,''
Astropart. Phys. \textbf{131}, 102605 (2021)
doi:10.1016/j.astropartphys.2021.102605
[arXiv:2008.11284 [astro-ph.CO]];
%\bibitem{DiValentino:2020vvd}
%E.~Di Valentino, L.~A.~Anchordoqui, \"O.~Akarsu, Y.~Ali-Haimoud, L.~Amendola, N.~Arendse, M.~Asgari, M.~Ballardini, S.~Basilakos and E.~Battistelli, \textit{et al.}
%``Cosmology Intertwined III: $f \sigma_8$ and $S_8$,''
Astropart. Phys. \textbf{131}, 102604 (2021)
doi:10.1016/j.astropartphys.2021.102604
[arXiv:2008.11285 [astro-ph.CO]];
%\bibitem{DiValentino:2020srs}
%E.~Di Valentino, L.~A.~Anchordoqui, \"O.~Akarsu, Y.~Ali-Haimoud, L.~Amendola, N.~Arendse, M.~Asgari, M.~Ballardini, S.~Basilakos and E.~Battistelli, \textit{et al.}
%``Snowmass2021 - Letter of interest cosmology intertwined IV: The age of the universe and its curvature,''
Astropart. Phys. \textbf{131}, 102607 (2021)
doi:10.1016/j.astropartphys.2021.102607
[arXiv:2008.11286 [astro-ph.CO]].

%F(R) gravity papers

\bibitem{Capozziello:2002rd}
S.~Capozziello,
%``Curvature quintessence,''
Int. J. Mod. Phys. D \textbf{11}, 483-492 (2002) doi:10.1142/S0218271802002025
[arXiv:gr-qc/0201033 [gr-qc]];
%\bibitem{Nojiri:2003ft}
S.~Nojiri and S.~D.~Odintsov,
%``Modified gravity with negative and positive powers of the curvature: Unification of the inflation and of the cosmic acceleration,''
Phys. Rev. D \textbf{68}, 123512 (2003) doi:10.1103/PhysRevD.68.123512
[arXiv:hep-th/0307288 [hep-th]];
%\bibitem{Nojiri:2006gh}
S.~Nojiri and S.~D.~Odintsov,
%``Modified f(R) gravity consistent with realistic cosmology: From matter dominated epoch to dark energy universe,''
Phys. Rev. D \textbf{74}, 086005 (2006) doi:10.1103/PhysRevD.74.086005
[arXiv:hep-th/0608008 [hep-th]];
%\bibitem{SaezGomez:2008uj}
D.~Saez-Gomez,
%``Modified f(R) gravity from scalar-tensor theory and inhomogeneous EoS dark energy,''
Gen. Rel. Grav. \textbf{41}, 1527-1538 (2009) doi:10.1007/s10714-008-0724-3
[arXiv:0809.1311 [hep-th]];
%\bibitem{Elizalde:2009gx}
E.~Elizalde and D.~Saez-Gomez,
%``F(R) cosmology in presence of a phantom fluid and its scalar-tensor counterpart: Towards a unified precision model of the universe evolution,''
Phys. Rev. D \textbf{80}, 044030 (2009) doi:10.1103/PhysRevD.80.044030 [arXiv:0903.2732
[hep-th]];
%\bibitem{Goheer:2009ss}
N.~Goheer, J.~Larena and P.~K.~S.~Dunsby,
%``Power-law cosmic expansion in f(R) gravity models,''
Phys. Rev. D \textbf{80}, 061301 (2009) doi:10.1103/PhysRevD.80.061301 [arXiv:0906.3860
[gr-qc]];
%\bibitem{OdintsovOF:2020}
  S. D. Odintsov, V. K. Oikonomou and F. P. Fronimos,
%f(R) Gravity kkk-Essence Late-time Phenomenology
 Phys. Dark  Univ. 29 (2020) 100563,   arXiv:2004.08884/ % [gr-qc]

\bibitem{Starobinsky:1980te}
%A.~A.~Starobinsky,
%``A New Type of Isotropic Cosmological Models Without Singularity,''
%Adv. Ser. Astrophys. Cosmol. \textbf{3}, 130-133 (1987) doi:10.1016/0370-2693(80)90670-X
%4869 citations counted in INSPIRE as of 06 Nov 2020
%\bibitem{Bamba:2014wda}
K.~Bamba, S.~Nojiri, S.~D.~Odintsov and D.~S\'aez-G\'omez,
%``Inflationary universe from perfect fluid and $F(R)$ gravity and its comparison with observational data,''
Phys. Rev. D \textbf{90} (2014) 124061, %doi:10.1103/PhysRevD.90.124061
[arXiv:1410.3993 [hep-th]].

\bibitem{Nojiri:2007as}
S.~Nojiri and S.~D.~Odintsov,
%``Unifying inflation with LambdaCDM epoch in modified f(R) gravity consistent with Solar System tests,''
Phys. Lett. B \textbf{657}, 238-245 (2007) doi:10.1016/j.physletb.2007.10.027
[arXiv:0707.1941 [hep-th]].

\bibitem{Nojiri:2005pu}
S.~Nojiri and S.~D.~Odintsov,
%``Unifying phantom inflation with late-time acceleration: Scalar phantom-non-phantom transition model and generalized holographic dark energy,''
Gen. Rel. Grav. \textbf{38}, 1285-1304 (2006) doi:10.1007/s10714-006-0301-6
[arXiv:hep-th/0506212 [hep-th]];
%\bibitem{Nojiri:2009kx}
S.~Nojiri, S.~D.~Odintsov and D.~Saez-Gomez,
%``Cosmological reconstruction of realistic modified F(R) gravities,''
Phys. Lett. B \textbf{681}, 74-80 (2009) doi:10.1016/j.physletb.2009.09.045
[arXiv:0908.1269 [hep-th]].
%\bibitem{CognolaENOSZ08}
G. Cognola, E. Elizalde, S. Nojiri, S. D. Odintsov, L. Sebastiani and S. Zerbini, Phys.
Rev. D 77 (2008) 046009, arXiv:0712.4017. % [hep-th].



\bibitem{HuSawicki07}
 W. Hu and I. Sawicki, %Models of f(R) Cosmic Acceleration that Evade Solar-System Tests,
Phys. Rev. D 76 (2007) 064004, arXiv:0705.1158.

\bibitem{delaCruz-Dombriz:2015tye}
\'A.~de la Cruz-Dombriz, P.~K.~S.~Dunsby, S.~Kandhai and D.~S\'aez-G\'omez,
%``Theoretical and observational constraints of viable f(R) theories of gravity,''
Phys. Rev. D \textbf{93}, no.8, 084016 (2016) doi:10.1103/PhysRevD.93.084016
[arXiv:1511.00102 [gr-qc]].

\bibitem{Arbey:2021gdg}
A.~Arbey and F.~Mahmoudi,
%``Dark matter and the early Universe: a review,''
Prog. Part. Nucl. Phys. \textbf{119}, 103865 (2021)
doi:10.1016/j.ppnp.2021.103865
[arXiv:2104.11488 [hep-ph]].

%Axions

\bibitem{Marsh:2015xka}
D.~J.~E.~Marsh,
%``Axion Cosmology,''
Phys. Rept. \textbf{643}, 1-79 (2016)
doi:10.1016/j.physrep.2016.06.005
[arXiv:1510.07633 [astro-ph.CO]].
%\bibitem{Sikivie:2006ni}
P.~Sikivie,
%``Axion Cosmology,''
Lect. Notes Phys. \textbf{741}, 19-50 (2008)
doi:10.1007/978-3-540-73518-2\_2
[arXiv:astro-ph/0610440 [astro-ph]].

\bibitem{Marsh:2017yvc}
M.~C.~D.~Marsh, H.~R.~Russell, A.~C.~Fabian, B.~P.~McNamara, P.~Nulsen and C.~S.~Reynolds,
%``A New Bound on Axion-Like Particles,''
JCAP \textbf{12}, 036 (2017)
doi:10.1088/1475-7516/2017/12/036
[arXiv:1703.07354 [hep-ph]].
%\bibitem{Irastorza:2018dyq}
I.~G.~Irastorza and J.~Redondo,
%``New experimental approaches in the search for axion-like particles,''
Prog. Part. Nucl. Phys. \textbf{102}, 89-159 (2018)
doi:10.1016/j.ppnp.2018.05.003
[arXiv:1801.08127 [hep-ph]].
%\bibitem{CAST:2017uph}
V.~Anastassopoulos \textit{et al.} [CAST],
%``New CAST Limit on the Axion-Photon Interaction,''
Nature Phys. \textbf{13}, 584-590 (2017)
doi:10.1038/nphys4109
[arXiv:1705.02290 [hep-ex]].

\bibitem{Caputo:2019tms}
A.~Caputo, L.~Sberna, M.~Frias, D.~Blas, P.~Pani, L.~Shao and W.~Yan,
%``Constraints on millicharged dark matter and axionlike particles from timing of radio waves,''
Phys. Rev. D \textbf{100}, no.6, 063515 (2019)
doi:10.1103/PhysRevD.100.063515
[arXiv:1902.02695 [astro-ph.CO]].

\bibitem{Soda:2017dsu}
J.~Soda and Y.~Urakawa,
%``Cosmological imprints of string axions in plateau,''
Eur. Phys. J. C \textbf{78}, no.9, 779 (2018)
doi:10.1140/epjc/s10052-018-6246-6
[arXiv:1710.00305 [astro-ph.CO]].

\bibitem{Cardoso:2018tly}
V.~Cardoso, \'O.~J.~C.~Dias, G.~S.~Hartnett, M.~Middleton, P.~Pani and J.~E.~Santos,
%``Constraining the mass of dark photons and axion-like particles through black-hole superradiance,''
JCAP \textbf{03}, 043 (2018)
doi:10.1088/1475-7516/2018/03/043
[arXiv:1801.01420 [gr-qc]].
%\bibitem{Day:2019bbh}
F.~V.~Day and J.~I.~McDonald,
%``Axion superradiance in rotating neutron stars,''
JCAP \textbf{10}, 051 (2019)
doi:10.1088/1475-7516/2019/10/051
[arXiv:1904.08341 [hep-ph]].
%\bibitem{Baryakhtar:2020gao}
M.~Baryakhtar, M.~Galanis, R.~Lasenby and O.~Simon,
%``Black hole superradiance of self-interacting scalar fields,''
Phys. Rev. D \textbf{103}, no.9, 095019 (2021)
doi:10.1103/PhysRevD.103.095019
[arXiv:2011.11646 [hep-ph]].

\bibitem{Oikonomou:2022tux}
V.~K.~Oikonomou,
%``Kinetic axion F(R) gravity inflation,''
Phys. Rev. D \textbf{106}, no.4, 044041 (2022)
doi:10.1103/PhysRevD.106.044041
[arXiv:2208.05544 [gr-qc]].

\bibitem{Oikonomou:2023bah}
V.~K.~Oikonomou,
%``Effects of the axion through the Higgs portal on primordial gravitational waves during the electroweak breaking,''
Phys. Rev. D \textbf{107}, no.6, 064071 (2023)
doi:10.1103/PhysRevD.107.064071
[arXiv:2303.05889 [hep-ph]].


%Exponential f(R) gravity

\bibitem{BambaGL:2010}
%%
K. Bamba, C. Q. Geng and C. C. Lee, %Cosmological evolution in exponential gravity,
J. Cosmol. Astropart. Phys. 08 (2010) 021,  arXiv:1005.4574.



\bibitem{ElizaldeNOSZ11}
 E. Elizalde, S. Nojiri, S.D. Odintsov, L. Sebastiani and S. Zerbini,
Phys. Rev. D. 83 (2011) 086006, arXiv:1012.2280.

\bibitem{OdintsovSGS:2017}
 S.~D. Odintsov, D. Saez-Chillon Gomez, G.~S. Sharov,
% Is exponential gravity a viable description for the whole cosmological history?
 Eur.\ Phys.\ J.\ C { 77} (2017) 862, arXiv:1709.06800.

 \bibitem{OdintsovSGSFlog:2019}
 S.~D. Odintsov, D. Saez-Chillon Gomez and G.~S. Sharov,
% Testing logarithmic corrections on $R^2$-exponential gravity by observational data
 Phys. Rev. D. { 99} (2019) 024003, arXiv:1807.02163.


%F(R) and axions

\bibitem{OdintsovOik_UniAx:2019}
 S. D. Odintsov and  V. K. Oikonomou,
% Unification of Inflation with Dark Energy in f(R) Gravity and Axion Dark Matter
Phys. Rev. D. 99 (2019) 104070,
 arXiv:1905.03496.  % [gr-qc]

\bibitem{OdintsovOik_Axion:2020}
 S. D. Odintsov and  V. K. Oikonomou,
% Geometric Inflation and Dark Energy with Axion $F(R)$ Gravity
Phys. Rev. D. 101 (2020) 044009,
 arXiv:2001.06830.  % [gr-qc]

%\bibitem{OdintsovOik_Ax_EPL:2020}
% S. D. Odintsov and V. K. Oikonomou,
%% Aspects of Axion F(R) Gravity
% EPL 129 (2020) 4, 40001 %doi:10.1209/0295-5075/129/40001
%arXiv:2003.06671. %[gr-qc]

\bibitem{Oikonomou_Uni:2021}
 V. K. Oikonomou,
 %Unifying inlfation with early and late dark energy epochs in axion F(R) gravity,"
 Phys. Rev. D 103 (2021)
044036, arXiv:2012.00586. % [astro-ph.CO].

\bibitem{OikonomouFTR:2023}
 V. K. Oikonomou, F. P. Fronimos, P. Tsyba and O. Razina,
% Kinetic Axion Dark Matter in String Corrected f(R) Gravity
 Phys. Dark Univ. 40 (2023) 101186,
 arXiv:2302.07147.  % [gr-qc]

\bibitem{Planck13}
Planck Collaboration, P. A. R. Ade { et al.} % {\it Planck 2013 results. XVI. Cosmological parameters.}
 Astron. Astrophys. 571 (2014) A16, arXiv:1303.5076.
%\bibitem{Planck15}
%Planck Collaboration, P. A. R. Ade { et al.} % {\it Planck 2015 results. XIII. Cosmological parameters.}
%Astron. Astrophys. 594 (2016) A13,
% arXiv:1502.01589. % [astro-ph.CO].
 \bibitem{Planck18}
Planck Collaboration,  N. Aghanim  { et al.}  {\it Planck 2018 results. VI. Cosmological
parameters.} Astron. Astrophys.  641 (2020) A6,
{arXiv:1807.06209}. % [astro-ph.CO].

% \bibitem{NojiriOSGS:2021}
%S.Nojiri, S.D. Odintsov, D. Saez-Gomez, G.S. Sharov,
%% Modelling and testing the equation of state for (Early) dark energy
% Phys. Dark Univ. 32 (2021) 100837, arXiv:2103.05304. % [gr-qc], 10.1016/j.dark.2021.100837
%




%\bibitem{Sharov:2016}
%G. S. Sharov,
%%{\it Observational constraints on cosmological models with Chaplygin gas and quadratic equation of state.}
%J. Cosmol. Astropart. Phys. {06} (2016) 023, arXiv:1506.05246.
%
%\bibitem{PanSh:2017}
%S. Pan and G. S. Sharov, %{\it A model with interaction of dark components and recent observational data.}
% Mon. Not. Roy. Astron. Soc. {472}  (2017)  4736, arXiv:1609.02287.
%
%\bibitem{HST2022}
%A. G. Riess,  W. Yuan, L. M. Macri, D. Scolnic, D. Brout  et al., %
%%A Comprehensive Measurement of the Local Value of the Hubble Constant with 1 km s-1 Mpc-1
%% Uncertainty from the Hubble Space Telescope and the SH0ES Team
%Astrophys. J. Lett. 934 (2022) 1, L7, {arXiv:2112.04510}. % [astro-ph.CO]


\bibitem{Scolnic17}
D. M. Scolnic  et al., % {\it The Complete Light-curve Sample of Spectroscopically Confirmed Type Ia Supernovae
% from Pan-STARRS1 and Cosmological Constraints from The Combined Pantheon Sample}
Astrophys. J. {859} (2018) 101, arXiv:1710.00845.

%\bibitem{HuangWW2015}
% Huang Q.-G., Wang K. and Wang S.  {\it Distance Priors from Planck 2015 data.}
%J. Cosmol. Astropart. Phys. 2015, \textbf{1512}, 022, }{arXiv: 1509.00969})

 \bibitem{OdintsovSGS:2022}
 S.~D. Odintsov, D. Saez-Chillon Gomez and G.~S. Sharov,
% Testing  viable extensions of Einstein-Gauss-Bonnet gravity
 Phys. Dark Univ. 37 (2022) 101100, arXiv:2207.08513.

  \bibitem{OdintsovOS:2023}
 S.~D. Odintsov, V. K.Oikonomou and G.~S. Sharov,
% Early dark energy with power-law F(R) gravity
 Physics Lett. B. 843 (2023) 137988,  arXiv:2305.17513.

 %H(z) data

 \bibitem{HzData}
 J. Simon, L. Verde and R. Jimenez, Phys. Rev. D 71 (2005) 123001, astro-ph/0412269; D. Stern, R. Jimenez, L. Verde,
M. Kamionkowski and S. A. Stanford, JCAP 1002 (2010) 008, arXiv:0907.3149; M. Moresco et
al., JCAP 1208 (2012) 006, arXiv:1201.3609; C. Zhang et al., Res. Astron. Astrophys. 14
(2014) 1221, arXiv:1207.4541; M. Moresco, Mon. Not. Roy. Astron. Soc. 450(1) (2015) L16,
arXiv:1503.01116; M. Moresco et al., JCAP 1605 (2016) 014, arXiv:1601.01701; A. L.
Ratsimbazafy et al. Mon. Not. Roy. Astron. Soc. 467(3) (2017) 3239, arXiv:1702.00418; N.
Borghi, M. Moresco, A. Cimatti,
% Towards a Better Understanding of Cosmic Chronometers: A new measurement of $H(z)$ at $z= 0.7$.
Astrophys. J. Lett. 928 (2022) 1, L4, arXiv:2110.04304.


\bibitem{ChenHuangW2018}
 L. Chen, Q.-G. Huang and K. Wang,
 % {\it Distance priors from Planck final release.}
J. Cosmol. Astropart. Phys. {1902} (2019) 028, arXiv:1808.05724.

\bibitem{Percival:2009}
W.~J.~Percival, B. A. Reid, D. J. Eisenstein, N. A. Bahcall, T. Budavari, \textit{et
al.} [SDSS],
%``Baryon Acoustic Oscillations in the Sloan Digital Sky Survey Data Release 7 Galaxy Sample,''
Mon. Not. Roy. Astron. Soc. \textbf{401} (2010), 2148, % -2168% doi:10.1111/j.1365-2966.2009.15812.x
[arXiv:0907.1660]. % [astro-ph.CO]]. %1432 citations counted in INSPIRE as of 25 Feb 2021

\bibitem{Blake:2011}
C.~Blake, E.~Kazin, F.~Beutler, T.~Davis, D.~Parkinson, S.~Brough, M.~Colless,
C.~Contreras, W.~Couch and S.~Croom, \textit{et al.}
%``The WiggleZ dark energy Survey: mapping the distance-redshift relation with baryon acoustic oscillations,''
Mon. Not. Roy. Astron. Soc. \textbf{418} (2011), 1707, %-1724 %doi:10.1111/j.1365-2966.2011.19592.x
[arXiv:1108.2635]. % [astro-ph.CO]]. %743 citations counted in INSPIRE as of 25 Feb 2021

\bibitem{Akaike}
 H. Akaike, IEEE Transactions on Automatic Control, 19 (1974) 716.

\bibitem{HuSugiyama95}
 W. Hu and N. Sugiyama,
% Small Scale Cosmological Perturbations: An Analytic Approach
  Astrophys. J. {471} (1996) 542, %-570,
 [arXiv:astro-ph/9510117]. % [astro-ph]].

\bibitem{Kazin:2009}
E.~A.~Kazin, M. R. Blanton, R. Scoccimarro, C. K. McBride, A. A. Berlind, \textit{et
al.}
%The Baryonic Acoustic Feature and Large-Scale Clustering in the SDSS LRG Sample
Astrophys. J. 710 (2010), 1444,%-1461
[arXiv:0908.2598].

\bibitem{Beutler:2011}
F.~Beutler, C.~Blake, M.~Colless, D.~H.~Jones, L.~Staveley-Smith, L.~Campbell,
Q.~Parker, W.~Saunders and F.~Watson,
%``The 6dF Galaxy Survey: Baryon Acoustic Oscillations and the Local Hubble Constant,''
Mon. Not. Roy. Astron. Soc. \textbf{416} (2011), 3017. %-3032 %doi:10.1111/j.1365-2966.2011.19250.x
[arXiv:1106.3366]. % [astro-ph.CO]]. %1542 citations counted in INSPIRE as of 25 Feb 2021

\bibitem{Chuang:2013}
C.~H.~Chuang, F.~Prada, A.~J.~Cuesta, D.~J.~Eisenstein, E.~Kazin, N.~Padmanabhan,
A.~G.~Sanchez, X.~Xu, F.~Beutler and M.~Manera, \textit{et al.}
%``The clustering of galaxies in the SDSS-III Baryon Oscillation Spectroscopic Survey: single-probe measurements and the strong power of normalized growth rate on constraining dark energy,''
Mon. Not. Roy. Astron. Soc. \textbf{433} (2013), 3559, % doi:10.1093/mnras/stt988,
[arXiv:1303.4486]. % [astro-ph.CO]].

\bibitem{Anderson:2013}
L.~Anderson \textit{et al.} [BOSS],
%``The clustering of galaxies in the SDSS-III Baryon Oscillation Spectroscopic Survey: baryon acoustic oscillations in the Data Releases 10 and 11 Galaxy samples,''
Mon. Not. Roy. Astron. Soc. \textbf{441} (2014) no.1, 24, %-62 doi:10.1093/mnras/stu523
[arXiv:1312.4877]. % [astro-ph.CO]]. %1102 citations counted in INSPIRE as of 25 Feb 2021

\bibitem{Ross:2014}
A.~J.~Ross, L.~Samushia, C.~Howlett, W.~J.~Percival, A.~Burden and M.~Manera,
%``The clustering of the SDSS DR7 main Galaxy sample \textendash{} I. A 4 per cent distance measure at $z = 0.15$,''
Mon. Not. Roy. Astron. Soc. \textbf{449} (2015) no.1, 835, %-847 doi:10.1093/mnras/stv154
[arXiv:1409.3242]. % [astro-ph.CO]]. %860 citations counted in INSPIRE as of 25 Feb 2021

\bibitem{Beutler:2016}
F.~Beutler, C.~Blake, M.~Colless, D.~H.~Jones, L.~Staveley-Smith, L.~Campbell,
Q.~Parker, W.~Saunders and F.~Watson,
%``The 6dF Galaxy Survey: Baryon Acoustic Oscillations and the Local Hubble Constant,''
Mon. Not. Roy. Astron. Soc. \textbf{464} (2017) 3, 3409,%-3430
[arXiv:1607.03149].

\bibitem{Chuang:2017}
C.~H.~Chuang, H.-J. Seo, A. J. Ross, P. McDonald, S. Saito, \textit{et al.}
%``The Clustering of Galaxies in the Completed SDSS-III Baryon Oscillation Spectroscopic Survey:
% Baryon Acoustic Oscillations in Fourier-space''
Mon. Not. Roy. Astron. Soc. \textbf{471} (2017) 2, 2370, %-2390,
[arXiv:1607.03151].

\bibitem{Bourboux:2017}
H. du Mas des Bourboux, J.-M. Le Goff, M. Blomqvist, N. G. Busca, J. Guy, J. Rich,
\textit{et al.}
% Baryon acoustic oscillations from the complete SDSS-III  Lya-quasar cross-correlation function at z = 2.4
 Astron. Astrophys. 608 (2017) A130, [arXiv:1708.02225].

\bibitem{Zhu:2018}
F. Zhu, N. Padmanabhan, A. J. Ross, M. White, W. J. Percival, \textit{et al.}
%The clustering of the SDSS-IV extended Baryon Oscillation Spectroscopic Survey DR14 quasar sample:
% Measuring the anisotropic Baryon Acoustic Oscillations with redshift weights
Mon. Not. Roy. Astron. Soc. 480 (2018) 1, 1096,%-1105
[arXiv:1801.03038].

\bibitem{Blomqvist:2019}
M. Blomqvist, H. du Mas des Bourboux, N. G. Busca, V. de Sainte Agathe, J. Rich,
%Baryon acoustic oscillations from the cross-correlation of Lya absorption and quasars in eBOSS DR14
Astron. Astrophys. 629 (2019) A86, [arXiv:1904.03430].

\bibitem{Hou:2020}
J. Hou, A. G. Sanchez, A. J. Ross, A. Smith, R. Neveux, \textit{et al.}
%The Completed SDSS-IV extended Baryon Oscillation Spectroscopic Survey: BAO and RSD measurements from anisotropic clustering analysis of the Quasar
%Sample in configuration space between redshift 0.8 and 2.2
Mon. Not. Roy. Astron. Soc. 500 (2020) 1, 1201,%-1221
[arXiv:2007.08998].

\bibitem{Tamone:2020}
A. Tamone, A. Raichoor, C. Zhao, A. de Mattia, C. Gorgoni, \textit{et al.}
% The Completed SDSS-IV extended Baryon Oscillation Spectroscopic Survey: Growth rate of structure
% measurement from anisotropic clustering analysis in configuration space between redshift 0.6 and 1.1 for the Emission Line Galaxy sample
Mon. Not. Roy. Astron. Soc. 499 (2020) 4, 5527,%-5546
[arXiv:2007.09009].


\bibitem{Vagnozzi:2022moj}
S.~Vagnozzi, R.~Roy, Y.~D.~Tsai, L.~Visinelli, M.~Afrin, A.~Allahyari, P.~Bambhaniya, D.~Dey, S.~G.~Ghosh and P.~S.~Joshi, \textit{et al.}
%``Horizon-scale tests of gravity theories and fundamental physics from the Event Horizon Telescope image of Sagittarius A,''
Class. Quant. Grav. \textbf{40}, no.16, 165007 (2023)
doi:10.1088/1361-6382/acd97b
[arXiv:2205.07787 [gr-qc]].


\end{thebibliography}
\end{document}